\documentclass[twocolumn]{revtex4}  
\usepackage[draft]{hyperref} 
\usepackage{amsfonts}
\usepackage{graphicx}        

\newcommand\eqnref[1]{(\ref{#1})}
\newcommand\figref[1]{Fig.~\ref{#1}}
\newcommand\tableref[1]{Table~\ref{#1}}

\newcommand{\iu}   {\mathrm{i}}     

\newcommand{\etal}   {\emph{et al. }}

\begin{document}

\title{Quasi-Homogeneous Backward-Wave Plasmonic Structures:\\
Theory and Accurate Simulation}

\author{Igor Tsukerman}
\affiliation{Department of Electrical and Computer Engineering, The
University of Akron, OH 44325-3904, USA}

\email{igor@uakron.edu}

\begin{abstract}
Backward waves and negative refraction are shown to exist in
plasmonic crystals whose lattice cell size is a very small fraction
of the vacuum wavelength (less than 1/40th in an illustrative
example). Such ``quasi-homogeneity''  is important, in particular,
for high-resolution imaging. Real and complex Bloch bands are
computed using the recently developed finite-difference calculus of
``Flexible Local Approximation MEthods'' (FLAME) that produces
linear eigenproblems, as opposed to quadratic or nonlinear ones
typical for other techniques. FLAME dramatically improves the
accuracy by incorporating local analytical approximations of the
solution into the numerical scheme.
%
\end{abstract}



\maketitle

%
Backward waves (Poynting vector opposite to phase velocity
\footnote{More generally, at an obtuse angle; but cases of strong
anisotropy \cite{Hoffman07} are beyond the scope of this paper.})
and the closely related phenomenon of negative refraction have been
extensively studied in recent years (e.g. \cite{Pendry04,Shalaev06}
and references there) due to the intriguing physical effects and
potential applications in imaging and other areas.

Backward waves in periodic structures may in general exist only if
the lattice cell size, as a fraction of the vacuum wavelength
$\lambda_0$, is above certain thresholds derived recently in
\cite{Tsukerman-JOSAB-08}. Plasmonic crystals
\cite{Shvets-PRL04,Davanco07} are an interesting exception because
in the vicinity of a plasmon resonance the constraints on the cell
size are relaxed or removed, as explained below, and thus it may be
possible to reduce the lattice cell size and approach an ideal
homogeneous negative-index medium.

Bloch modes play a central role in the analysis of electromagnetic
waves in periodic structures. While analytical expressions for these
modes are available only in special one-dimensional cases
\cite{Yeh05,Sakoda05,Tsukerman-book07}, there exist a variety of
computational techniques: Fourier transforms (plane wave expansion,
PWE) \cite{Johnson01,Sakoda05,Tsukerman-book07}, scattering matrices
and lattice summation \cite{Botten05}, finite difference
\cite{Yang96,Yu04} and finite element
\cite{Mias99,Axmann99,Dobson01,Davanco07,Tavallaee08} analysis,
semi-analytical methods \cite{Yuan06}, and more.

Bloch wave problems have three (in 2D) or four (in 3D) scalar
eigenparameters: frequency $\omega$ and the Cartesian components of
the Bloch vector $\mathbf{K}$. Solving for all these parameters, and
the respective eigenmodes, simultaneously is impractical. The usual
approach is to look for the values of $\omega$ for any given
$\mathbf{K}$. The differential operator of the problem, and hence
the respective matrices in the numerical computation, contain the
permittivity $\epsilon$ and therefore for dispersive media
($\epsilon = \epsilon(\omega)$) depend on the frequency in a
complicated way. Consequently, the resulting eigenvalue problems
with respect to $\omega$ are nonlinear. Several solution methods
have been proposed \cite{Toader04,Spence05} but are not simple, and
convergence is not guaranteed.

A more elegant approach, where frequency is treated as a given
parameter and components of the Bloch vector as unknown eigenvalues,
has been explored relatively recently in PWE \cite{Shi05} and in FEM
\cite{Tavallaee08,Davanco07}. \emph{Quadratic} eigenproblems with
respect to the Bloch number usually arise and can be converted to
linear ones by introducing auxiliary unknowns either on the
continuous level (e.g. solving for both fields $\mathbf{E}$,
$\mathbf{H}$ instead of just one) or, alternatively, on the linear
algebra level \cite{Tisseur01}. This conversion doubles the number
of unknowns; the computational cost, typically proportional to the
cube of the system size, increases about eightfold.

In the recently developed generalized finite-difference (FD)
calculus of Flexible Local Approximation MEthods (FLAME
\cite{Tsukerman06,Tsukerman-book07,Tsukerman-PBG08}) high accuracy
is achieved by replacing the Taylor expansions of standard FD
analysis with much better approximating functions, e.g. plane waves
or cylindrical harmonics. In FLAME, $\omega$ is a natural
``independent variable'' because the approximating functions are
derived for a fixed value of $\omega$. FLAME has two clear
advantages in the computation of (real or complex) Bloch modes: (i)
it dramatically improves the accuracy by incorporating local
analytical approximations of the solution into the numerical scheme
(see examples below); (ii) it produces linear eigenproblems.

Let us consider band structure calculation in a photonic crystal
formed by an infinite lattice of rectangular cells $L_x \times L_y$
in the $xy$-plane. In a very common case, each cell contains a
dielectric cylindrical rod with a radius $r_\mathrm{rod}$ and the
relative dielectric permittivity $\epsilon_\mathrm{rod}$. The medium
outside the rod has permittivity $\epsilon_\mathrm{out}$. At optical
frequencies, all media are intrinsically nonmagnetic. The governing
wave equation for the TE mode (one-component magnetic field phasor
$H = H_z$) is
\vskip -0.25in
\begin{equation}\label{eqn:wave-eqn-Hz}
    \nabla \cdot \epsilon^{-1} \nabla H ~+~ \omega^2 \mu_0 H ~=~ 0
\end{equation}
\vskip -0.1in \noindent where the relative dielectric permittivity
$\epsilon = \epsilon(x, y)$ is periodic over the lattice. The
$\exp(-\iu \omega t)$ convention is used for complex phasors. The
$H$ field is sought as a Bloch-Floquet wave \cite{Sakoda05} with a
(yet undetermined) Bloch vector $\mathbf{K}_B$:
\vskip -0.25in
\begin{equation}\label{eqn:H-field-Bloch-wave}
    H(\mathbf{r}) ~=~ H_\mathrm{PER}(\mathbf{r}) \,
    \exp(\iu \mathbf{K}_B \cdot \mathbf{r})
    ~~~~ \mathbf{r} \equiv (x, y)
\end{equation}
\vskip -0.1in
\noindent where $H_\mathrm{PER}$ is periodic over the lattice. In
the space of Bloch vectors $\mathbf{K}_B$, the first Brillouin zone
is $[-\pi/L_x, \pi/L_x] \, \times \, [-\pi/L_y, \pi/L_y]$. For
notational simplicity and without real loss of generality, let $L_x
= L_y = a$.

\begin{figure}
  \centering
  \includegraphics[width=\linewidth]{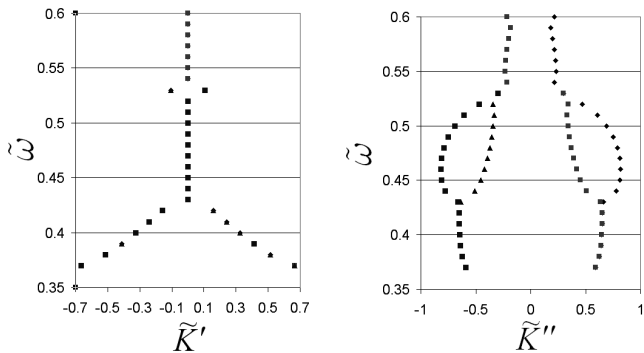}\\
  \caption{A fragment of the band diagram of the Davanco \etal
  \cite{Davanco07} plasmonic crystal. $\tilde{\omega} = \omega a / c$,
  $\tilde{K} = K_Ba / \pi$.}\label{fig:band-fragment-Davanco-crystal}
\end{figure}

There are two general options: solving for the periodic factor
$H_\mathrm{PER}(x, y)$ or, alternatively, for the full $H$-field of
\eqnref{eqn:wave-eqn-Hz}. In the first case, standard periodic
boundary conditions apply, but the differential operator is more
complicated than in the second case. The boundary conditions for the
full $H$-field are ``scaled-periodic'' due to the Bloch exponential
$\exp(\iu \mathbf{K} \cdot \mathbf{r})$:
%
\vskip -0.25in
\begin{equation}\label{eqn:scaled-periodic-bc-x-E-field}
    H \left( a/2, \, y \right) = \exp(\iu K_x a) \,
    H \left( -a/2, \, y \right); ~~
    |y| \leq a/2
\end{equation}
\vskip -0.1in \noindent with similar conditions at the boundaries $y
= \pm a/2$.

Accurate local analytical approximations that FLAME relies on are
available for the full $H$-field formulation and involve Bessel /
Hankel functions \cite{Tsukerman05,Tsukerman06,Tsukerman-book07}.
%
%
Two Bloch conditions -- one for the $H$ field and another one for
the $E$ field -- need to be imposed on the cell boundaries; the
implementation details are described in \cite{Tsukerman-PBG08}. In
matrix-vector form, the FLAME eigenvalue problem is
%
$
    L \underline{E} \,=\, (b_x B_x + b_y B_y) \underline{E}
$,
%
where $\underline{E}$ is the Euclidean vector of the nodal values of
the field on the grid, and $b_x$, $b_y$ are the Bloch factors
%
$
    b_x = \exp (\iu K_x L_x); ~
    b_y = \exp (\iu K_y L_y)
$.
%
Matrix $L$ has a sparse structure typical of finite-difference
methods; matrix $B$ is extremely sparse -- its nonzero entries
correspond only to two layers of grid nodes adjacent to the cell
boundary.

Numerical examples for real-$\mathbf{K}$ modes in regular
dielectrics are given in \cite{Tsukerman-PBG08}. To illustrate the
computation of \emph{complex} modes in plasmonic crystals, let us
start, for convenience of comparison, with the same example as in
\cite{Davanco07}. The permittivity of the rods is assumed to be
described by the normalized Drude model
%
$
     \epsilon(\tilde{\omega}) = 1 - \tilde{\omega}^{-1} (\tilde{\omega} - \iu
     \tilde{\omega}_c)^{-1}
$
%
with the normalized frequencies $\tilde{\omega} = \omega /
\omega_p$, $\tilde{\omega}_c = \omega_c / \omega_p$, where
$\omega_p$ is the plasma frequency, $\omega_c$ is the damping rate,
and $c$ is the speed of light in free space. The square lattice cell
size is $a = c/ \omega_p$ and the cylinder radii are $r_\mathrm{rod}
= 0.45 a$. The Bloch modes computed with the 9-point FLAME scheme
coincide with the ones reported in \cite{Davanco07}; a fragment of
the band diagram computed with FLAME for $\omega_c = 0$ is shown in
\figref{fig:band-fragment-Davanco-crystal} for reference. The
accuracy of FLAME is remarkably high: as evidenced by the typical
convergence results for several modes in
\tableref{table:convergence-Bloch-wavenumbers}, on the $40 \times
40$ grid the FLAME results already have about five correct digits --
several orders of magnitude higher accuracy than in a typical PWE
calculation \cite{Sakoda05,Tsukerman-book07,Tsukerman-PBG08}. Note
that this example is \emph{not} computationally favorable because
the filling factor is high and the gaps between the rods are narrow.

\begin{figure}
  \centering
  \includegraphics[width=\linewidth]{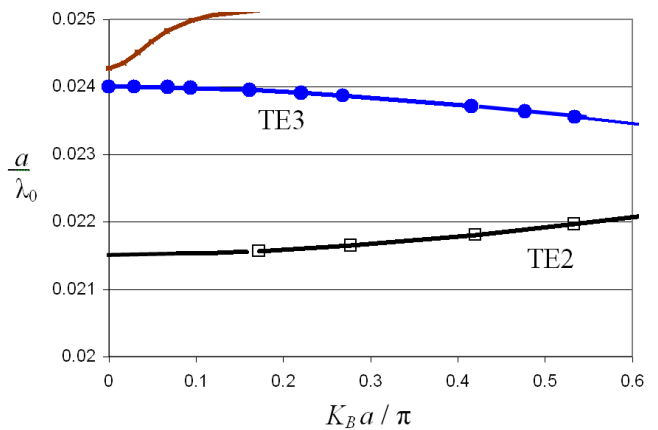}\\
  \caption{A backward-wave mode (TE3, circles, dotted line) in a plasmonic crystal with the
  lattice cell size as small as $\sim \lambda_0 / 40$.
  Parameters: $\tilde{\omega}_p = 0.25$, $r_\mathrm{rod} = 0.45a$.}\label{fig:bands-2-3-PlC-small-cell}
\end{figure}

\begin{table}
  \centering\begin{tabular}{|c|c|c|c|c|}
  \hline
   Grid ~&~ $K_\mathrm{real}$ ~&~ $K''_1$ ~&~ $K''_2$\\
   \hline   $20 \times 20$ ~&~ 0.241048852 ~&~ 2.573701526  ~&~  4.113530131 \\
   \hline   $30 \times 30$ ~&~ 0.240708525 ~&~ 2.572916824 ~&~ 4.108636747 \\
   \hline   $40 \times 40$ ~&~ 0.240652136 ~&~ 2.572740104 ~&~ 4.544368769 \\
   \hline   $50 \times 50$ ~&~ 0.240644493 ~&~ 2.572711175 ~&~ 4.544040773 \\
   \hline
   \end{tabular}
  \caption{Typical convergence of the numerical values for Bloch wavenumbers.
  Results for $\tilde{\omega} = 0.26$. $K_\mathrm{real}$ is a real mode.
  Modes $K''_{1,2}$ have the real part $K' = \pi/a$.
  }\label{table:convergence-Bloch-wavenumbers}
\end{table}

There is some similarity between FLAME and the semi-analytical
construction of \cite{Yuan06}, where the field \emph{in the whole
lattice cell} is expanded into cylindrical harmonics and an
eigenvalue problem for the coefficients of this expansion is
obtained by imposing the Bloch boundary conditions at a set of
collocation points on the cell boundary. In contrast, FLAME uses
only local analytical approximations that are valid over each
individual grid stencil. Therefore FLAME can be extended to more
general situations, e.g. with more than one rod in the cell,
non-circular shapes, defects, photonic waveguides, etc.
\cite{Pinheiro07,Dai-adaptive-FLAME08,Tsukerman-book07}, where
global analytical solutions are not available.

\begin{figure}
  \centering
  \includegraphics[width=\linewidth]{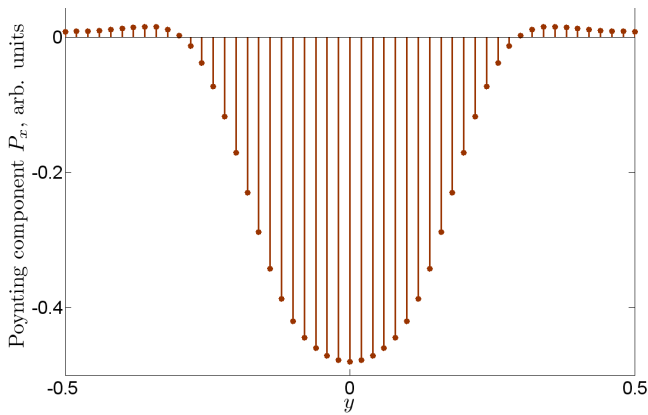}\\
  \caption{Predominantly negative power density on the right edge
  of the very small lattice cell
  ($a / \lambda_0 = 0.02387$).}
  \label{fig:Poynting-right-edge}
\end{figure}

\textbf{Backward waves}. As shown in \cite{Tsukerman-JOSAB-08}, the
cell size of periodic dielectric structures capable of supporting
backward waves must lie above a fundamental threshold:
\vskip -0.25in
\begin{equation}\label{eqn:eta-leq-4pi2-epsmax}
    | \eta | ~\leq~ (4\pi^2)^{-1} |\epsilon|_{\max} \left(1
    \,+\, |\lambda|_{\max} (\mathcal{L}_{\epsilon}^{-1})
    |\epsilon|_{\max} \right)
\end{equation}
\vskip -0.1in
\noindent where $\eta = \tilde{\omega}^{-2}$;  $\tilde{\omega} =
\omega a / c = 2\pi a / \lambda_0$; $|\lambda|_{\max}
(\mathcal{L}_{\epsilon}^{-1})$ is the maximum eigenvalue
of the inverse of the electrostatic operator $\mathcal{L}_{\epsilon}
= \tilde{\nabla} \cdot \epsilon \tilde{\nabla}$. This eigenvalue is
bounded unless the operating frequency is close to the quasi-static
plasmon resonance value. In particular, for non-plasmonic materials
with $0 < \epsilon_{\min} \leq \epsilon \leq \epsilon_{\max}$
throughout the lattice cell, the lower cell size bound specializes
to
%
%
$
    ( a / \lambda_0 )^2 \geq [ |\epsilon|_{\max} (1
    + |\epsilon|_{\max} / (4\pi^2 \epsilon_{\min}) ) ]^{-1}
$ \cite{Tsukerman-JOSAB-08}.
The plasmonic case is thus exceptional, because at the plasmonic
resonance the electrostatic operator in
\eqnref{eqn:eta-leq-4pi2-epsmax} becomes singular and the constraint
on the cell size disappears.


The following example demonstrates that backward waves, and hence
negative refraction, can be obtained in plasmonic crystals with very
small lattice cells (here $\sim \lambda_0 / 40$). As before,
consider a square lattice of cylindrical rods, with $r_\mathrm{rod}
= 0.45a$. For this ``proof-of-concept'' example, losses are
neglected and the Drude-like dielectric function is set as
$\epsilon_\mathrm{rod} = 1 - \tilde{\omega}_p^2 / \tilde{\omega}^2$,
with $\tilde{\omega}_p = \omega_p a / c$.
%
%
A fragment of the band diagram in the $\Gamma X$ direction for
$\tilde{\omega}_p = 0.25$ is shown in
\figref{fig:bands-2-3-PlC-small-cell}. The TE3 band (circles, solid
line) corresponds to a backward wave: (a) the group velocity
$\partial \omega / \partial K_B$ for this band is negative; (b) the
$x$ component of the Poynting vector, plotted in
\figref{fig:Poynting-right-edge}, clearly indicates negative energy
flow; (c)  the first-Brillouin-zone component of the wave is
dominant, as can be confirmed by Fourier analysis of the periodic
field $H_\mathrm{PER}(x, y)$ in \eqnref{eqn:H-field-Bloch-wave};
hence the wave has a well defined positive phase velocity.

In summary, negative refraction can be obtained in a
quasi-homogeneous medium with a very small cell size as a
``plasmonic exception'' circumventing the theoretical cell size
bounds of \cite{Tsukerman-JOSAB-08}. This may have important
implications for imaging, as the inhomogeneity of the medium limits
the resolution \cite{Smith03,Tsukerman-JOSAB-08}. For illustration,
a backward-wave mode is demonstrated in a plasmonic crystal with the
cell size of less than 1/40th of the vacuum wavelength. FLAME, a
generalized finite-difference calculus based on very accurate local
approximations of the solution, was applied to the computation of
real and complex Bloch bands.

\bibliographystyle{plain}


\end{document}